\documentclass[letterpaper]{JHEP3}
\usepackage{bm}

\newcommand{\be}{\begin{equation}}
\newcommand{\ee}{\end{equation}}
\newcommand{\bea}{\begin{eqnarray}}
\newcommand{\eea}{\end{eqnarray}}

\renewcommand{\b}[1]{\bar{#1}}

\newcommand{\del}{\partial}

\renewcommand{\ap}{\alpha^\prime}

\newcommand{\ot}{\overline{\Theta}}

\newcommand{\comment}[1]{}

\title{String Theoretic Bounds on Lorentz-Violating Warped Compactification}
\author{Andrew R. Frey\\ Department of Physics\\ University of California\\
Santa Barbara, CA 93106 USA\\ E-mail: \email{frey@vulcan.physics.ucsb.edu}}
\preprint{hep-th/0301189}
\abstract{We consider warped compactifications that solve the 10 dimensional
supergravity equations of motion at a point, stabilize the position of
a D3-brane world, and admit a warp factor that violates Lorentz invariance
along the brane.  This gives a 
string embedding of ``asymmetrically warped'' models which we use to calculate
stringy ($\ap$) corrections to standard model dispersion relations, paying
attention to the maximum speeds for different particles.  We find, from
the dispersion relations, limits on gravitational Lorentz violation in these
models, improving on current limits on the speed of graviton propagation,
including those derived from field theoretic loops.
We comment on the viability of models that use asymmetric warping for
self-tuning of the brane cosmological constant.}

\begin{document}

\section{Introduction}\label{s:intro}

Violation of Lorentz invariance is of great interest in the field of quantum
gravity since it can arise in many different ways \cite{Kostelecky:2001xz}.  
For example, canonical quantum gravity has been argued to predict 
Planck scale Lorentz violation \cite{Gambini:2001ia},
and noncommutative space-time coordinates break Lorentz invariance
\cite{Carroll:2001ws}.  Lorentz violation has also been related to 
variation of coupling constants \cite{Kostelecky:2002ca}.
The literature is
very broad, and we refer the reader to \cite{Kostelecky:2001xz} for a review
of several developments.

In this paper, we consider a form of Lorentz violation that can appear 
in braneworld compactifications of string theory, which is known as
asymmetric warping.  In braneworld compactifications, the spacetime manifold
need not be a direct product of noncompact and compact submanifolds; the
noncompact spacetime metric usually has a warp factor dependent on the 
internal space (see \cite{Randall:1999ee,Randall:1999vf,Rubakov:2001kp} for
early work on warping and a review).  However, nothing in the higher 
dimensional theory keeps the warp factors of the different noncompact 
dimensions the same, as first discussed in \cite{Kraus:1999it}.  
The metric takes the form 
\be\label{metric}
ds^2 = -e^{A(y)} dt^2 +e^{B(y)} d\vec x^2 + \tilde g_{mn}(y) dy^m dy^n\ .
\ee
A nice
feature of this form of Lorentz violation is that it is classically confined
to the gravitational sector (the graviton moves at a different speed than
light); if the brane stays at a fixed position in the
extra dimensions, we can rescale the noncompact coordinates and recover
an $SO(3,1)$ symmetry for fields confined to the brane, including the
standard model (in most braneworlds).  These models therefore avoid, at tree
level, the rather stringent limits on Lorentz violation in the standard model 
(see \cite{Jacobson:2002hd} for a recent review and \S\ref{s:limits} for
a brief discussion of the experimental bounds).  

Despite the classical restriction of Lorentz violation to gravity, we 
expect that a symmetry broken anywhere is broken everywhere.  Indeed, 
the communication of Lorentz violation from gravity to the standard model
by quantum loops was studied in \cite{Burgess:2002tb} in a rather general
framework.  Limits on Lorentz violation in the standard model are thus 
related to limits on asymmetric warping, which \cite{Burgess:2002tb} quoted
in terms of the difference of graviton and photon speeds.  Evading limits
such as these is a new problem for extra-dimensional cosmology, similar
to the original flatness problem that was solved by inflation 
\cite{Chung:2000ji}.

In parallel with \cite{Burgess:2002tb}, we consider the communication of 
Lorentz violation from gravity to the standard model sector in a string
theoretical background, which we describe in section \ref{s:background}.  
Instead of loop corrections, we consider $\ap$ corrections to the D3-brane
action, which are at tree level in string perturbation theory.  Since
the worldsheet disk fluctuates in the dimensions orthogonal to the D-brane,
these terms are sensitive to the Lorentz violating derivatives of the metric.
(Using stringy corrections has the slight advantage over loops that we do
not have to worry about how to regulate perturbative gravity.)  We
calculate and describe the Lorentz violating contributions (from 
geometry only) to standard 
model propagators in section \ref{s:corrections}.  Because the $\ap$ 
corrections we calculate involve only the metric, they do not give all 
possible terms, but they give estimates that are independent of the background
of the other supergravity fields. In section \ref{s:limits},
we then use our results to put limits on Lorentz violation in 
Randall-Sundrum \cite{Randall:1999ee,Randall:1999vf} type braneworlds and
comment on asymmetric warping in self-tuning braneworlds
\cite{Csaki:2000dm,Csaki:2001mn}.  Our limits imply that the energy density
that is self-tuned cannot be within a few orders of magnitude of the higher
dimensional fundamental scale, so there must be some other mechanism to
explain the smallness of the brane energy density.  We close with a
short discussion in section \ref{s:discussion}.  We do not address a solution
to the flatness problem of \cite{Chung:2000ji}, but our results do sharpen
it somewhat.

\section{A Simple String Braneworld}\label{s:background}

In this section, we describe a braneworld model that solves the 10D type IIB
supergravity equations of motion locally at the position of the brane.
We treat the brane as a string theoretic D3-brane in the approximation that
it probes the ambient geometry without affecting it.  Since we are interested
in the effect of the geometry on the D3-brane fields, we justify the 
probe approximation by noting that a static brane should not be affected
by its own action on the background (as in electrostatics).  We will choose
a background that stabilizes the position of the D3-brane and allows for
asymmetrical warping.  We are most
interested in making order of magnitude estimates for Lorentz-breaking
effects, we will not try to incorporate the standard model.

For simplicity, we take the type IIB 3-form field strengths and RR scalar 
to vanish and dilaton to be constant with string coupling $g_s$.  The
equations of motion for the 5-form field strength and the string frame
metric then become (see \cite{Polchinski:2000uf}, for example, using
the conventions of \cite{Polchinski:1998rr})
\bea
R_{MN} &=& \frac{g_s^2}{96}\tilde F_{MPQRS}\tilde F_N{}^{PQRS}
\label{einstein}\\
d\tilde F = 0 &,& \star\tilde F = \tilde F \ .\label{5form}
\eea
We want to confine Lorentz violation to the gravitational sector, so we can
take an ansatz\footnote{We take Greek indices on spacetime, lower case 
Roman from the middle of the alphabet on the internal manifold, 
and capital Roman for all 10 dimensions.  We will usually use vector notation
for spatial directions of the noncompact spacetime but will sometimes 
denote them with indices $i,j,k$. Hats will indicate orthonormal basis 
indices, as for a tangent space.}
\bea
ds^2 &=& -e^{A(y)} dt^2 +e^{B(y)} d\vec x^2 + \tilde g_{mn}(y) dy^m dy^n
\nonumber\\
\tilde F_{\mu\nu\lambda\rho m}&=& -\frac{1}{g_s} \epsilon_{\mu\nu\lambda\rho}
\del_m F(y)\ \ ,\ \ 
\tilde F_{mnpqr} = \frac{1}{g_s}\epsilon_{mnpqr}{}^s \del_s F(y)\ 
,\label{ansatz}
\eea
where $\epsilon$ is the volume form on spacetime or the internal 
manifold respectively.  Note that this ansatz automatically satisfies the
5-form equations (\ref{5form}).
At the position of the D3-brane, $y^m = 0$, the
$y^m$ are Riemann normal on the internal manifold\footnote{Through the rest
of this paper, we use ``Riemann normal coordinates'' to refer to coordinates
that are Riemann normal on the 6D internal manifold, ignoring the rest of the
metric.} ($\tilde g_{mn}(y=0)=
\delta_{mn}$, $\del_p\tilde g_{mn}(y=0)=0$), 
and the functions $A,B,F$ have Taylor expansions
$A(y) = a_m y^m + (1/2)a_{mn} y^m y^n +\cdots$, etc.

At zeroth order in $y^m$, the Einstein equation (\ref{einstein}) becomes
\bea
R_{00}&=& \frac{1}{2}\left(a_m a^m +a_m{}^m \right)
= \frac{1}{4} f_mf^m\nonumber\\
R_{ij}&=& \frac{1}{2}\left(b_m b^m +b_m{}^m \right)\delta_{ij} 
= -\frac{1}{4} f_m f^m 
\delta_{ij}\label{long0}
\eea
along the D3-brane and
\be\label{transverse}
R_{mn}=\tilde R_{mn} +\frac{1}{2} a_{mn}-\frac{3}{2} b_{mn} = -\frac{1}{4}
\left( 2 f_m f_n - f_pf^p \delta_{mn}\right)
\ee
transverse to it, where $\tilde R_{mn}$ is the Ricci tensor for the internal
metric $\tilde g_{mn}(y)$.  The other components are trivial.  
Note that eqn (\ref{transverse}) is extraneous,
since we can satisfy it by choosing manifolds with the appropriate 
$\tilde R_{mn}$ (for example, many cases are satisfied with 
a warped Calabi-Yau manifold).  Since we will also want the second derivative
of the function $F$ later, we should look at the next order in $y^m$;
along the brane, we have
\bea
\left(a_n a^n +a_n{}^n\right)a_m +2 a^n a_{nm}&=&  f^n f_{mn}
\nonumber\\
\left(b_n b^n +b_n{}^n\right)b_m +2 b^n b_{nm}&=&  f^n f_{mn}
\ . \label{long1}
\eea

To work within perturbation theory, we should stabilize the D-brane position.
From the action for D-branes (see, for reviews, 
\cite{Polchinski:1998rr,Johnson:2000ch}), the potential for the brane
scalars -- the brane coordinates -- is the determinant of the pulled-back
metric minus the 4-form potential,
\be\label{potential} V = \frac{\tau}{g_s} e^{(A+3B)/2} - \tau C_{0123}
\ee
with brane tension $\tau/g_s$.
For the forces on the brane to balance, we require
\be\label{force}
\frac{1}{2}a_m +\frac{3}{2} b_m +f_m = 0\ .\ee
The scalar masses come from the second derivative of the potential,
\be\label{masssqr}
\frac{2g_s}{\tau} \del_m\del_n V = \frac{1}{2} \left( a_m+3b_m\right)
\left( a_n +3b_n\right) +a_{mn}+3b_{mn} +2 g_s f_{mn}\ .\ee
To stabilize all the brane coordinates, the trace of this matrix must be
positive.  Using eqns (\ref{long0},\ref{force}), we get
\be\label{trace}
\frac{2g_s}{\tau} \del^2 V = f_mf^m +2g_s f_m{}^m -a_ma^m 
-3b_mb^m\ .\ee
The equations of motion (\ref{long1}) seem not to constrain $f_{mn}$ tightly,
so there seems no obstacle to stabilizing the brane position.  We should 
note, however, that without the 5-form, the potential is at best flat or is
at a saddle point.  (In fact, with $a_m=b_m$, we get just the $f_m{}^m$ term.)

We close this section by noting that this is by no means the most general
way to stabilize the brane position; for example,
a nontrivial dilaton and 3-form flux
can contribute to the scalar potenial for multiple D-branes 
\cite{Myers:1999ps,Johnson:2000ch} 
(\cite{Grana:2002nq} gives a detailed mapping of the SUGRA fields to MSSM
parameters in a particular model).  However, since our purpose is to 
demonstrate that $\ap$ corrections communicate Lorentz breaking from the
pure gravity sector to the brane fields, our background is sufficient.
From this point forward, we need only refer to the geometrical part of
our background (\ref{ansatz}).

\section{Stringy Tree Level Corrections}\label{s:corrections}

In this section, we will identify terms in the action for brane fields
that violate Lorentz invariance due to worldsheet interactions with the
background metric.  We begin by examining the zero-th order action for
all the brane fields and verifying that there are no violations of Lorentz
invariance.  Then we move to leading order $\ap$ corrections for the 
scalars and gauge fields.  Finally, we comment on corrections we do not
calculate but which should appear.  Through all the analysis, we look for
terms at second order in the brane fields because most experimental limits
come from Lorentz violating dispersion relations.  The strongest limits
are on different effective ``speeds of light'' (more appropriately,
maximum attainable velocities, or MAVs)
for different fields, so we will calculate just the difference of MAVs from
one for various D-brane fields.  

\subsection{$\kappa$ Symmetric D-brane Action}\label{ss:kdbi}

Including fermions, the action for a D3-brane 
\cite{Cederwall:1997ri,Bergshoeff:1997tu} is
\begin{equation}
S=-\tau\int d^4\zeta\,  e^{- \bm{\Phi}}\sqrt{-\det \left(
\bm{g}_{\alpha\beta} 
+ \bm{\mathcal{F}}_{\alpha\beta}\right)}+\tau\int 
e^{\bm{\mathcal{F}}} \wedge\bm{C}\ .
\label{susyaction}
\end{equation}
The fields in boldface are superfields in type II 10D superspace, where
\begin{equation}
\bm{\mathcal{F}}=2\pi\ap F-\bm{B}\ ,\ \mathnormal{and}\ 
\bm{C}=\oplus_n \bm{C}_{(n)}
\label{somedefs}\end{equation}
is the collection of the pullbacks of the RR potentials.  Here, Greek
letters from the beginning of the alphabet refer to directions tangent
to the D3-brane, which are to lowest order in the brane fluctuations
$\zeta^\alpha = \delta^\alpha_\mu x^\mu$.  There is a local $\kappa$
symmetry on the worldvolume that reduces the Grassman superspace coordinates
to a single $SO(9,1)$ Majorana-Weyl spinor $\Theta$ that contains four
worldvolume $SO(3,1)$ fermions. 

The expansion of this action in terms of the D-brane scalars $Y^m$, 
gauge field $F_{\alpha\beta}$, and fermions $\Theta$ was worked out
in \cite{Grana:2002tu} for a background similar to ours (slightly more 
general but Lorentz invariant).  It is straightforward to write down
the renormalizable part of the action for our background (with the
flat metric at $y^m=0$):
\bea
S&=& -\frac{\tau}{g_s}\int d^4x \left[ \frac{1}{2} \delta_{mn}
\del_\mu Y^m \del^\mu Y^n +\frac{g_s}{\tau} V(Y) +\frac{(2\pi \ap)^2}{4}
F_{\mu\nu}F^{\mu\nu} \right.\nonumber\\
&&\left. -i \ot\Gamma^\mu \del_\mu \Theta +\frac{i}{4} 
\omega_{\mu\hat M\hat N} \ot \Gamma^\mu \Gamma^{\hat M\hat N}\Theta
\right]\ .\label{kdbi}
\eea
The tension is given by $\tau = (2\pi)^{-1}(2\pi \ap)^{-2}$\footnote{This
gives a gauge coupling of $g_{YM}^2=g_s$.  Since we are taking a single D3,
we ignore terms associated with nonAbelian gauge theory.} 
and the potential
should only be taken out to quartic order in $Y^m$ (there are additional
terms at higher order in $Y,\Theta$ 
if we include nonrenormalizable interactions).  The last term includes
$\omega_{\mu\hat M\hat N}$, the 10D spin connection 1-form in the direction
along the brane.  The only vanishing components are
\be\label{spincon}
\omega_0{}^{\hat 0 \hat m}=-\frac{1}{2} a_m e^{m \hat m} \ ,
\ \ \omega_i^{\hat j \hat m} = -\frac{1}{2}b_m e^{m\hat m} \delta^{\hat j}_i
\ ,\ee
so the final term becomes
\be\label{lastterm}
-\frac{i}{8} \left(a_m+3b_m\right) \ot \Gamma^m\Theta \ .\ee
It has been shown in \cite{Grana:2002tu} that this linear combination of
spinors vanishes due to the Majorana-Weyl nature of $\Theta$.  At $Y^m=0$,
this is the only term that could have communicated the Lorentz breaking
from the gravity sector to the brane fields, so we must turn to stringy
corrections.\footnote{Clearly, there are Lorentz breaking terms in the
nonrenormalizable part, since the metric $g_{\mu\nu}$ is not Lorentz
invariant at $y^m \neq 0$.  We want instead changes to dispersion
relations, since limits on those are known.}

\subsection{$(\ap)^2$ Corrections to Dispersion Relations}\label{ss:ap2}

Even at tree level in string amplitudes, the action (\ref{susyaction}) 
is incomplete; integrating out higher string modes gives a series of higher
derivative terms with $\ap$ as the expansion parameter.  We see below
that, when some of the derivatives act on the background, these corrections
can actually modify the propagator (in the language of loops, renormalize
the wavefunction).  Next, we will review the $\mathcal{O}(\ap)^2$ corrections 
that we consider, along with the necessary geometry.

\subsubsection{Corrected DBI Action and Embedding Geometry}\label{sss:embed}

Probably the most famous $\ap$ corrections to D-brane actions are from the
couplings to the RR potentials; Riemann curvature becomes lower-dimensional
brane charge.  The Wess-Zumino part of the action is
\be\label{aroof}
S_{WZ}=
\tau \int C\wedge e^{\mathcal{F}} \wedge \sqrt{\frac{\hat A\left( 4\pi^2
\ap R_T\right)}{\hat A \left(4\pi^2 \ap R_N\right)}}\ee
where $\hat A$ is the Dirac ``A-roof genus'' and $R_T,R_N$ are the tangent
and normal bundle Riemann tensors to be defined below
(see \cite{Cheung:1998az} for a full derivation and
\cite{Johnson:2000ch} for a review).  
Although these corrections
are generally nonvanishing for our background, we are not interested in 
them.  We do note that they modify the background, since now the D3-brane 
will have charge under the RR scalar, but we this is a subleading effect
and will not affect our conclusions.

To get the modified dispersion relations, we need to consider the corrections
to the DBI action, which have been studied in \cite{Bachas:1999um,
Wyllard:2000qe,Fotopoulos:2001pt,Wyllard:2001ye,Barabanschikov:2003fr}.  
The most complete 
results for curved backgrounds are in \cite{Fotopoulos:2001pt}, but only 
the geometry is considered and not other SUGRA fields.  The DBI action becomes
(considering the bosonic part only)
\bea
S_{DBI}&=& -\tau\int d^4\zeta\,  e^{- \Phi}\sqrt{-\det \left(
g_{\alpha\beta} 
+ \mathcal{F}_{\alpha\beta}\right)} \left[ 1 -\frac{(2\pi\ap)^2}{192}
\left( (R_T)_{\alpha\beta\gamma\delta}(R_T)^{\alpha\beta\gamma\delta}  
\right.\right.\nonumber\\
&&\left.\left. -2(R_T)_{\alpha\beta}(R_T)^{\alpha\beta}
-(R_N)_{\alpha\beta \hat a \hat b}(R_N)^{\alpha\beta \hat a \hat b} 
+2\b R_{\hat a \hat b}\b R^{\hat a \hat b}\right) \right]\label{corrected}
\eea
up to $\mathcal{O}(\ap)^2$.  There is an additional contribution at this 
order with an undetermined coefficient, but it vanishes on-shell, so it does
not affect S-matrix elements or dispersion relations \cite{Fotopoulos:2001pt}.
Therefore, we can ignore it.  Here, $\hat a, \hat b$ are normal bundle 
indices in an orthonormal basis with vielbein $\xi^{\hat a}$.

We now define the various Riemann and Ricci tensors above, as is discussed
in \cite{Bachas:1999um,Wyllard:2000qe,Fotopoulos:2001pt,Wyllard:2001ye}.
In the following, we use $P[\cdots]$ to denote the pullback to the 
worldvolume or pushforward to the normal bundle,
but for brevity we write $g_{\alpha\beta}\equiv P[g]_{\alpha\beta}$ and
$g^{\alpha\beta} \equiv (P[g]_{\alpha\beta})^{-1}$.
Start with the extrinsic curvature, or second fundamental form, of the
D-brane embedding
\be\label{extrinsic}
\Omega^{\hat a}_{\alpha\beta} = \xi^{\hat a}_M \left( \del_\alpha\del_\beta
X^M -(\Gamma_T)^\gamma_{\alpha\beta}\del_\gamma X^M +
\Gamma^M_{NP}\del_\alpha X^N\del_\beta X^P\right)\ ,\ee
where $(\Gamma_T)^\gamma_{\alpha\beta}$ is the Christoffel connection of
the pulled-back metric.  The tangent and normal bundle Riemann tensors 
can be shown to be
\bea
(R_T)_{\alpha\beta\gamma\delta}&=& P[R]_{\alpha\beta\gamma\delta}+
\delta_{\hat a\hat b} \left( \Omega^{\hat a}_{\alpha\gamma}
\Omega^{\hat b}_{\beta\delta}-\Omega^{\hat a}_{\alpha\delta}
\Omega^{\hat b}_{\beta\gamma}\right)\ ,\label{riemannt}\\
(R_N)_{\alpha\beta}{}^{\hat a \hat b} &=& P[R]_{\alpha\beta}{}^{\hat a\hat b}
+g^{\gamma\delta}\left( \Omega^{\hat a}_{\alpha\gamma}
\Omega^{\hat b}_{\beta\delta} -\Omega^{\hat b}_{\alpha\gamma}
\Omega^{\hat a}_{\beta\delta}\right)\ .\label{riemannn}\eea
Then $(R_T)_{\alpha\beta}$ is just the Ricci tensor associated with the 
tangent Riemann tensor, and
\be\label{rbar}
\b R^{\hat a\hat b} = g^{\alpha\beta}P[R]_\alpha{}^{\hat a\hat b}{}_\beta
+g^{\alpha\gamma}g^{\beta\delta}\Omega^{\hat a}_{\alpha\beta}
\Omega^{\hat b}_{\gamma\delta}\ .\ee

As we continue, we should remember that the action (\ref{corrected}) assumes
that all the other fields vanish.  We will address this point in 
\S\ref{ss:other}, and we will below use T-duality to deduce possible couplings
between the extrinsic curvature and the worldvolume field strength.  Also,
the terms involving fermions are not known, although 
\cite{Barabanschikov:2003fr,Howe:2001wc} 
give steps in that direction.  We will consider that the corrections to
fermions will be similar to those for the scalars, which we can calculate.

\subsubsection{Maximum Attainable Velocities for Scalars}\label{sss:mavs}

To get the modified MAVs for the scalar fields on the brane, we just need
to calculate the Lorentz violating kinetic terms that arise in the 
corrected action (\ref{corrected}).  We can write these all as
$(\cdots)\vec{\del}Y\cdot\vec{\del}Y$ (because $\del_0 Y\del_0 Y$ differs
by a Lorentz invariant term).  Knowing this allows us to simplify our
calculation greatly; we need keep only terms up to $\mathcal{O}(\del Y)^2$
and can ignore terms in which $Y$ appears without a derivative as well as
$\del^2 Y$ terms.  Also,
since the background would be Lorentz invariant if the warp factors
were equal, $A=B$, we can write $A=B+\Lambda$ and take only terms in which
$\Lambda$, the Lorentz violating function, appears.  To get the best limits
on Lorentz violation, we keep only terms that are linear in $\Lambda$.

The key to our results is that the extrinsic curvature has terms that are
zero-th order in the brane fluctuations that come from $\Gamma^m_{\mu\nu}$.  
In our background, the extrinsic curvature is
\bea
\Omega^m_{00} &=& \frac{1}{2} a^m -a_n\del_0 Y^n\del_0 Y^m+\cdots\ ,
\nonumber\\
\Omega^m_{0i} &=& -\frac{1}{2} a_n\del_i Y^n \del_0 Y^m -\frac{1}{2} b_n
\del_0 Y^n \del_i Y^m+\cdots\ ,\nonumber\\
\Omega^m_{ij} &=& -\frac{1}{2} b^m \delta_{ij} -b_n \del_{\left( i\right.}
Y^n\del_{\left. j\right)} Y^m+\cdots\ .\label{omega}
\eea
We use standard notation regarding symmetrization of indices with a weighting
of $1/2$.  Also, to reduce proliferation of indices, we replace the normal
bundle index $\hat a$ with the coordinate index $m$ since they are the
same at lowest order in perturbation theory.  We have still included all
the appropriate terms from the expansion of the normal bundle vielbein, 
however.  Additional zero-th order terms appear in the Riemann tensor part of
$\b R^{mn}$.  We leave those, along with the details of the rest of the
calculation to the appendix \ref{a:calc}.

After much algebra, and carefully accounting for all terms, we find the
following Lorentz violating kinetic terms for the scalars to linear order
in $\Lambda$:
\bea
\delta S &=& \frac{\tau}{g_s} \int d^4 x \frac{(2\pi\ap)^2}{192}
\vec{\del} Y^m\cdot \vec{\del} Y^n \left[ \frac{21}{2}b^2 b_{\left( m\right.}
\lambda_{\left. n\right)} +5 b\cdot \lambda b_m b_n -\frac{1}{2}b^2
\lambda_{mn}+b\cdot \lambda b_{mn}
-4\lambda^pb_{p\left(m\right.}b_{\left. n\right)}\right.\nonumber\\
&&\left. 
-4 b^pb_{p\left( m\right.}\lambda_{\left. n\right)} +\frac{47}{4} 
b^2 b\cdot \lambda
\delta_{mn} +4 b^{pq}\lambda_{pq}\delta_{mn}+4\lambda^{pq}b_p b_q
\delta_{mn}+20 b^{pq}\lambda_p b_q \delta_{mn}\right]\ .
\label{scalarmav}
\eea
Because the action (\ref{corrected}) does not include contributions from the
background 5-form, these are probably not the only Lorentz violating
kinetic terms for the scalars.  However, because Lorentz invariance is
broken by the background, there is no reason to suppose that the terms we
have calculated are precisely canceled by those that we have not.  Therefore,
we can take these as an estimate of the change in MAV for the scalars.
From (\ref{scalarmav}), we can see clearly why these terms modify the
MAV of the scalars; by a \textit{field dependent} redefinition of the
``speed of light,'' we can clearly combine all the time and space derivatives
of the fields into the usual form $\del_\mu Y\del^\mu Y$.  Each scalar
has a Lorentz invariant kinetic action, but the Lorentz groups are in 
general different for the different scalars.  This way of thinking about
our corrections is essentially the formalism of \cite{Coleman:1998ti}.

Of course, even at this order in the string expansion, there are many other 
corrections to the action.  For example, the constant terms in the 
extrinsic curvature give a cosmological constant at when the brane
is at $y^m=0$; more generally, there is a new contribution to the potential
for the D-brane position.  We are not really expanding around the vacuum of
the theory.  Similarly, the supergravity background should be altered (even
if we had incorporated the tension and charge of the brane to begin with)
because the A-roof corrections induce a charge for other supergravity 
fields, notably the RR scalar.  We will not worry about these effects,
since they will be suppressed by $\ap{}^2 b^4$, and the warp factor should be
somewhat less curved than the string scale for supergravity to be valid
in the first place.  Thus, in perturbation theory, they do not affect the 
leading Lorentz violation.  Also, as we mentioned above, there will be other 
corrections due to the 5-form field strength, and we know that for the 
``black 3-brane'' background
the corrections should all cancel because of supersymmetry.  As another
example, the scalar wavefunctions will be renormalized by Lorentz
invariant kinetic terms in the correction; again, this is suppressed by
$\ap{}^2 b^4$.  Finally, there are many, generally nonrenormalizable,
interactions introduced.

\subsubsection{Maximum Attainable Velocity for Photon}\label{sss:photon}

To find the communication of Lorentz violation to electrodynamics from
gravity, we proceed by a somewhat indirect route.  The derivative corrections
to the DBI action are not known in a curved background, so, rather than
generalize the flat space results of \cite{Wyllard:2000qe,Wyllard:2001ye},
we use T-duality to relate the field strength to the scalars.  We will
also argue that our results are plausible terms in the curved-space
generalization of \cite{Wyllard:2000qe,Wyllard:2001ye}.

Our T-duality argument runs as follows.  Suppose we compactify one of the
noncompact spatial dimensions, say $x^3$, on a large circle.   Then
we perform T-duality on $x^3$, giving a metric of 
\be\label{tdmetric}
ds^2 = -e^A dt^2 +e^B \sum_{i=1}^2 \left( dx^i\right)^2 +e^{-B}\left( dx^3
\right)^2 +\tilde g_{mn} dy^m dy^n \ee
with, as before, all components of the metric depending only on $y^m$.
Now $x^3$ is compactified on a small circle, and the braneworld is a D2-brane.
In fact, this metric is of the same form as our original metric (\ref{ansatz})
except that $x^3$ is not a Riemann normal coordinate.  The only difference 
this makes from the calculations of
\S\ref{sss:mavs} is that we now have to keep in mind that there is a 
nonvanishing Christoffel symbol $\Gamma^m_{33}$ that contributes kinetic terms
to $\Omega^m_{\alpha\beta}$ ($\Gamma^3_{3m}$ will not 
matter because $\Omega^3_{\alpha\beta}$ will already be second order in 
$\del X,\del Y$).  Otherwise some numerical factors differ.  If
we recalculate eqn (\ref{scalarmav}) for fluctuations of $X^3$ in 2 dimensions,
we find
\be\label{x3mav}
\delta S = \frac{\tau_2}{g_{s,2}} \int d^3x \frac{(2\pi\ap)^2}{192}
\left| \vec\del X^3 \right|^2 \left[ \frac{11}{4} b^2 b\cdot\lambda 
+3 b^{mn}\lambda_{mn} +3\lambda^{mn}b_mb_n +9b^{mn}\lambda_mb_n\right]
\ .\ee
Here 
$\tau_2,g_{s,2}$ are the appropriate tension and string coupling for the 
D2-brane case.  

If we then T-dualize back to the D3-brane and take the $x^3$ circle radius
to infinity, we get back the original tension and string coupling and take
$X^3\to (2\pi\ap)A_3$.  Then by gauge invariance, we must replace
$(2\pi\ap)\del_i A_3 \to \mathcal{F}_{i3}$, and isotropy requires that we
promote $\mathcal{F}_{i3}\mathcal{F}^{i3}\to \mathcal{F}_{ij}\mathcal{F}^{ij}$.
Therefore, we end up with a correction to the photon kinetic term of
\be\label{photmav}
\delta S = \frac{\tau}{g_{s}} \int d^4x \frac{(2\pi\ap)^2}{192}
\mathcal{F}_{ij}\mathcal{F}^{ij} \left[ \frac{11}{4} b^2 b\cdot\lambda 
+3 b^{mn}\lambda_{mn} +3\lambda^{mn}b_mb_n +9b^{mn}\lambda_mb_n\right]
\ .\ee
So we see that even the speed of light is modified; the MAV of a photon is
not 1!

Let us check this term against the known $\mathcal{O}(\ap)^2$ corrections
for the worldvolume gauge field.  For a flat background, \cite{Wyllard:2000qe}
showed that the corrections can be given in terms of a non-symmetric
metric $h_{\alpha\beta}= g_{\alpha\beta}+\mathcal{F}_{\alpha\beta}$ and
its Riemann tensor.  So in curved space, we could easily have 
$(\mathnormal{Riemann,Ricci})^2\mathcal{F}^2$ terms, including such terms
as $\b R_{mn}\Omega^m_{ik}\Omega^n_{jl}\mathcal{F}^{ij}\mathcal{F}^{kl}$.  
As it turns out,
the method of replacing $g_{\alpha\beta}\to h_{\alpha\beta}$ only works
to $\mathcal{O}(\ap)^2$ for the Wess-Zumino terms when the background is
curved, but we are working only to that order in any case.

\subsection{Other Corrections}\label{ss:other}

We already discussed some of the other corrections that appear for scalars
at the end of \S \ref{sss:mavs}.  These are $\mathcal{O}(\ap)^2$ terms that
we could have calculated but have not (because the experimental limits are
far greater on the dispersion relations that we have calculated).  One
additional correction of this type is of second order in the scalars but
fourth order in derivatives; they introduce quartic terms in the dispersion
relation.  Some of them T-dualize to a modification for the field strength
of the form $\vec\del\mathcal{F}_{ij}\cdot \vec\del\mathcal{F}^{ij}$, and
there are actually experimental 
limits on such terms.  However, they are very weak
compared to the limits on the second order terms \cite{Jacobson:2002hd}.

We are not just limited to $\mathcal{O}(\ap)^2$ corrections, even though
these are the only ones known (and incompletely, at that).  Because the 
extrinsic curvature and some of the Riemann components have terms that are
zero-th order in D-brane fields and their derivatives, we expect that all
orders in $\ap$ should contribute to kinetic terms, with a suppression
by the string length over the warp factor curvature length.  This 
should be small for supergravity to remain valid, but it should still be within
a few orders of magnitude of unity in a Randall-Sundrum type model with only
one fundamental scale \cite{Randall:1999ee,Randall:1999vf}.  Of interest
is whether the Riemann tensor of the internal metric $\tilde g_{mn}$ can
contribute to the Lorentz violating terms at some order in the $\ap$ 
expansion; in appendix \ref{a:calc}, we see that it couples to the scalars
only in a Lorentz invariant way in the terms we consider.  Since, in any
realistic (say a warped Calabi-Yau) compactification, we expect 
the curvature of the compact manifold to be of order the string scale (and
therefore the fundamental Planck scale), such terms could contribute to
very high orders in $\ap$.

We also mention again the fact that there should be corrections due to 
the other background bulk supergravity fields even at $\mathcal{O}(\ap)^2$.
Unless they also couple to the background metric, we would not expect them
to violate Lorentz invariance as long as their background does not, but
it would be interesting to be able to study them.

Most importantly, the worldvolume fermions should also have Lorentz violation
kinetic terms from $\ap$ corrections to the $\kappa$ symmetric D-brane action.
Unfortunately, the $\mathcal{O}(\ap)^2$ corrections to the DBI action are
not known in $\kappa$ symmetric form, so we cannot calculate them.  
Progress in this direction was reported in \cite{Howe:2001wc}, but we will
simply estimate that the Lorentz violating terms will be roughly the same
for the fermions as for the scalars.

We close this section with a comment about the applicability of our results, 
then.  We have calculated only 
the purely gravitational contribution to the Lorentz violating kinetic terms
and only at lowest order in $\ap$.  It is very likely, at least in some
models, that contributions from other fields or even from higher orders in
$\ap$ (because the internal Riemann curvature should go as $1/\ap$) could 
also be significant.  Therefore, the reader should not consider the limits
we derive here to be definitive for any specific model (especially since the
stringy details of most models are not yet resolved).  Rather, the limits
we get are somewhat rough but model independent because they depend only
on the metric.  Additionally, because the Lorentz invariance is broken by the
background geometry, extra contributions to the $\ap$ corrections should
not change the order of magnitude of our limits.  
Almost any model of asymmetric warping, if considered as a
solution of string theory, will be subject to them.

\section{Limits on Asymmetric Warping}\label{s:limits}

We can now use the results of \S\ref{s:corrections} to apply known 
experimental limits on Lorentz violation in the standard model to 
asymmetric warping in the gravitational sector.  In particular, many 
unobserved effects, such as \v{C}erenkov radiation by charged particles 
in vacuum and photon decay, can occur if the MAVs of photons and fermions
are different \cite{Coleman:1998ti}\footnote{We do not consider interactions,
such as electromagnetic muon decay, that also violate lepton number.}.  
In fact, vacuum \v{C}erenkov radiation and photon decay are so efficient 
that we should observe no particles above threshhold for those interactions
(see, for example \cite{Jacobson:2001tu}).  These two interactions give
the constraint that the magnitude of the difference of photon and electron
MAVs should not be greater than $10^{-16}$ (in units with the speed of light
equal to unity).  A demonstration of this result, along with a review of 
many dispersion relation tests, is given in \cite{Jacobson:2002hd}.

In fact, we could use more stringent constraints.  For example, ignoring the
parton structure, the proton MAV should differ from the photon MAV by no 
more than $10^{-22}$ \cite{Jacobson:2002hd}.  Atomic spectroscopy experiments
measuring spatial anisotropy of nuclear dipole and quadrupole couplings
give a similar (indirect) bound of $10^{-22}$ \cite{Lamoreaux:1986xz}; 
this is the limit used by \cite{Burgess:2002tb}.
Additionally, other Lorentz
(and even CPT) violating effects in QED have been considered by many authors
(see, for example, \cite{Kostelecky:2001mb,Bluhm:2001ms}).  However,
these mainly deal with polarization effects, and we will not consider them.
To be conservative, we will require that the difference in fermion and
photon MAVs be less than $10^{-16}$ in absolute value.  However, 
for comparison to the results of \cite{Burgess:2002tb}, we will also 
quote results using a limit of $10^{-22}$.

From equations (\ref{scalarmav},\ref{photmav}), we can therefore see that,
including the numerical factors, we should have
\be\label{bound}
\ap{}^2 (b_m)^3 \lambda_m , \ \ap{}^2 b_{mn}\lambda_m b_m ,\ 
\ap{}^2\lambda_{mn} (b_m)^2 ,\ 
\ap{}^2\lambda_{mn}b^{mn} < \epsilon \equiv 10^{-16},10^{-22} \ee
in terms of magnitude.
We will use this to put limits on the Lorentz violating function $\Lambda$
in the following subsection, and we will relate this bound to the 
speed of graviton propagation in the asymmetrically warped background.  
Like \cite{Burgess:2002tb}, we will find the range of parameter space when
our limits improve on the bound $|c_g-1|<10^{-6}$ (see 
\cite{Will:2001mx} for a discussion relating this bound to 
experiments).

\subsection{Perturbations around Randall-Sundrum}\label{ss:general}

The prototypical warped braneworld models are those of Randall and Sundrum
\cite{Randall:1999ee,Randall:1999vf} (see \cite{Rubakov:2001kp} for a 
general review), which have one or two branes in a total of 5 dimensions.  
Therefore, we take the asymmetric
warping to be a small perturbation around the Lorentz invariant 
Randall-Sundrum metric below.
To get our limits, we will assume that the other 5 dimensions of string
theory are compactified at the 5-dimensional Planck scale $M_5$, so the 
fundamental 10D scale, the string scale, is the 5D scale, $\ap = M_5^{-2}$.
It is trivial to extend our analysis to other cases given some specific
model and our results should not be significantly changed; 
however, we will focus on the most basic case here.  

If the visible sector brane is at $y=0$, the Randall-Sundrum metric is
\be\label{RSmetric}
ds^2 = e^{\mp 2 k |y|} dx^\mu dx_\mu + dy^2 \ee
where the -(+) sign corresponds to the one (two) brane model.  This is
$\mathnormal{AdS}_5$ with curvature given by the mass scale $k$.  To make the
supergravity approximation valid but to avoid naturalness problems, 
we should have $k \lesssim M_5$.  For concreteness, we will quote limits 
taking the somewhat arbitrary value $k = M_5/10$.

We should also briefly address the warping in the other 5 dimensions, since
they enter into the limits (\ref{bound}).  In a traditional compactification,
this warping vanishes, and, in more general warped compactifications, those
warp factors should not be larger than the fundamental Planck scale if we
can use a geometrical interpretation.  A related issue is the origin of the
5D cosmological constant.  In string theory, this would be related to 
the background supergravity fields, and, in the simple model of section
\ref{s:background}, the 5-form plays the role of a negative cosmological
constant (see the Einstein equations (\ref{long0},\ref{transverse})).  As
long as $f_m$ is aimed along the Randall-Sundrum $y$ direction, this is 
the same situation as in the $\mathnormal{AdS}_5\times S^5$ solution of
string theory.  Because of the equilibrium condition (\ref{force}), we would
not have large warp factors in the extra 5 dimensions without taking
$\lambda_m\sim b_m$.
Therefore, we will assume that the warping in the small compact dimensions
is no larger than $k$ and
does not affect the magnitude of our limits. This choice 
gives the most lenient limits in any event. 

Since the second derivatives of the warp factor vanish in the Randall-Sundrum
scenario, we get the limits
\be\label{RSlimits}
\frac{|\lambda_m|}{M_5} < \left(\frac{M_5}{k}\right)^3 \epsilon ,\ \
\frac{|\lambda_{mn}|}{M^2_5} < \left( \frac{M_5}{k}\right)^2 \epsilon\ .\ee
Putting in specific numbers for $k$ and $\epsilon$, we have
\be\label{RSlimits2}
\frac{|\lambda_m|}{M_5} < 10^{-13},10^{-19} ,\ \
\frac{|\lambda_{mn}|}{M^2_5} < 10^{-14},10^{-20}\ .\ee
These numbers suggest a type of hierarchy problem; perhaps the asymmetric
warping is caused by a black hole or other gravitating object very far from 
our brane in the $y$ direction.  This is the usual radius stabilization 
problem.

Let us now relate the limits (\ref{RSlimits},\ref{RSlimits2}) to the known
limits on the MAV for gravitons.  Specifically, we will ask in what region
of parameter space do our results improve the limit $|c_g - 1|<10^{-6}$.
To do so, we must relate the Lorentz violating function $\Lambda$ to 
the speed of gravitational wave propagation.  With a global solution, we
could do this by finding the higher dimensional zero-mode of the graviton,
but we will need to make do with perturbation theory since we have only
a solution expanded around $y^m=0$.  Our ansatz is to ignore the 5 small
dimensions and use the Randall-Sundrum zero-mode as the graviton wavefunction
in the warped dimension.  We use this wavefunction to get an expectation
value for the speed of gravity, $e^\Lambda$.

For perturbations around the one-brane model, we find that
\be\label{onebranecg}
c_g-1 \simeq \frac{\lambda_m}{M_5} \frac{M_5}{k},\ 
\frac{\lambda_{mn}}{M_5^2}\left(\frac{M_5}{k}\right)^2\ .\ee
Depending on which limit we use for the standard model MAVs, these give
us limits better than $|c_g - 1|<10^{-6}$ when $M_5/k < 10^{2.5}, 10^{4}$
for either $\lambda_m$ or $\lambda_{mn}$.  The two-brane model is
slightly more complicated because the expectation value depends on the
position of the regulator brane, $y_R$.  However, to solve the hierarchy
problem, typically $ky_R \simeq 10^2$.  Then we can approximate the graviton
speed by taking (\ref{onebranecg}) and replacing $k\to 1/y_R \simeq 10^{-2} k$.
This gives, for $\lambda_m$, an improvement over $|c_g - 1|<10^{-6}$ when
$M_5/k < 10^{2}, 10^{3.5}$ and, for $\lambda_{mn}$, an improvement when
$M_5/k < 10^{1.5}, 10^{3}$.  Since we expect $k\simeq M_5$, this covers
almost all of the expected parameter space.  We should also note that
there is a one-sided bound of $c_g - 1 > -10^{-15}$ from gravitational 
\v{C}erenkov radiation \cite{Moore:2001bv}.  Our bounds can compete with
or even improve on this more stringent bound (but without sign) when 
$k\simeq M_5$.  

We note briefly that these bounds compare well to the
field theoretic loop bounds of \cite{Burgess:2002tb}.  When using the
same bounds on standard model MAVs as \cite{Burgess:2002tb}, we find
an improvement on limits for $c_g$ in precisely the region of parameter
space that we naturally expect to occur.  This seems to have wider
applicability than bounds from loops, which are effective when the 
fundamental scale of the higher dimensions is itself small.  In warped
compactifications, this may not be the case.

\subsection{Comments on Cosmological Constant Self-Tuning}\label{ss:cc}

Now we will comment on the application of our results to models of 
asymmetric warping used to solve the cosmological constant problem.  As
is well known, the size of the cosmological constant is as yet unexplained
by theory.  The review \cite{Weinberg:1989cp} considers many different
possible resolutions of the cosmological constant problem, including 
the possibility of ``tuning'' by a scalar field that reduces the effect of 
the vacuum energy density.  The conclusion is that such a method does not
work in models with standard 4D physics.  Nonetheless, self-tuning has 
enjoyed a revival of interest with the discovery by 
\cite{Arkani-Hamed:2000eg,Kachru:2000hf} that vacuum energy density on a
brane can be translated into curvature of a bulk scalar.  Because of 
singularities in the bulk, however, these models require fine-tuning to
reproduce 4D gravity and
therefore do not excape the no-go theorem stated above 
\cite{Csaki:2000wz}.

Seemingly, the way around this difficulty is to place the singularity 
behind an event horizon; additionally, the spacetime metric can be written
in a 5D Schwarzschild-like form
\be\label{schwarzschild}
ds^2 = -h(r) dt^2 +(kr)^2 d\vec x^2 +h^{-1}(r) dr^2 \ ,\ee
where $k$ is the same as in the Randall-Sundrum case (for $h(r) = (kr)^2$,
this is just Randall-Sundrum in different coordinates).  These types of 
solutions with a charged black hole background were first studied in 
\cite{Kraus:1999it,Csaki:2000dm,Csaki:2001mn,Csaki:2001yz} and generalized in 
\cite{Grojean:2001pv,Nojiri:2001ae}.  

Even though the full metric does not have $SO(3,1)$ symmetry, 
\cite{Kraus:1999it,Csaki:2000dm,Csaki:2001mn,Csaki:2001yz} 
argue that these asymmetrically
warped models evade limits on standard model Lorentz violation because
brane fields feel an $SO(3,1)$ symmetry as long as the brane stays at a 
fixed position.  The only effect at the field theory tree level 
would be to alter the speed of gravitons.  We have seen that Lorentz violation
can be communicated to the standard model both by loops \cite{Burgess:2002tb}
and by $\ap$ corrections in string theory as in section \ref{s:corrections}.
It is natural to ask whether the limits we derived above are stringent
enough to rule out these models.

The difficulty is that the warp factors at the brane are determined entirely
by the energy density on the brane through jump conditions if we work only
in 5D \cite{Binetruy:1999ut}.  (See \cite{Binetruy:1999ut} for some of
the consequences this has for cosmology.)  Therefore, we find that the 
warp factors for Riemann normal coordinate $y$ are given by 
\be\label{ccwarping}
b_y = \frac{\rho}{3M_5^3},\ \lambda_y = -\frac{\rho}{M_5^3}
(1+\omega ),\ b_{yy}= -\frac{\rho^2}{6M_5^6}(1+\omega),\ 
\lambda_{yy} = 24k^2 +\frac{\rho^2}{54M_5^6}(7+60\omega),\ee
where $\rho$ is the brane energy density and 
the brane equation of state is $P=\omega \rho$.  To avoid a naked
singularity in the bulk, $\omega <-1$ \cite{Csaki:2000dm}.  Ignoring this
exotic equation of state and any other cosmological difficulties, our bounds
(\ref{bound}) become
\bea
\ap{}^2 b_y^3\lambda_y  
&\simeq& \left(\frac{\rho}{M_5^4} \right)^4 (1+\omega)
<\epsilon\nonumber \\
\ap{}^2 b_{yy} b_y \lambda_y &\simeq& \left(\frac{\rho}{M_5^4} \right)^4 
(1+\omega)^2 < \epsilon \nonumber\\
\ap{}^2 b_y^2 \lambda_{yy}  &\simeq& \left(\frac{k}{M_5}\right)^2
\left(\frac{\rho}{M_5^4}\right)^2 <\epsilon \nonumber\\
\ap{}^2 b_{yy}\lambda_{yy} &\simeq&  \left(\frac{k}{M_5}\right)^2
\left(\frac{\rho}{M_5^4}\right)^2 (1+\omega)<\epsilon\ .   \label{ccbounds}
\eea
assuming that $k/M_5$ is larger than $\rho/M_5^4$, which seems reasonable
if only because of power counting.  The best bound is the third one, since
it avoids ambiguity due to the unspecified value of $\omega$.  Taking
values for $k/M_5$ as before, we find $\rho/M_5^4 < 10^{-7},10^{-10}$.  
So we cannot use asymmetric warping to tune away 5D Planck scale energy
densities; some other mechanism must explain why the brane energy density
is at least a few orders of magnitude smaller than $M_5$.  This somewhat
lessens the appeal of the self-tuning mechanism.

It is worth noting that the constraints should be a bit tighter if we note that
the warping $B$ should not be determined solely by the brane energy density
when the setup is embedded into more than 5D (because the jump conditions
are no longer boundary conditions).  If we take, as seems likely in a 10D
scenario, $b_y \simeq k$, then we get
\be\label{ccbound2}
\left(\frac{k}{M_5}\right)^3 \frac{\rho}{M_5^4} (1+\omega) <\epsilon\ ,
\left(\frac{k}{M_5}\right)^4 <\epsilon\ .\ee
If we use $k/M_5\simeq 1/10$ as before, we find 
$\rho/M_5^4 (1+\omega)< 10^{-13},10^{-19}$ as in \S\ref{ss:general}.
However, 
the second relation gives $k/M_5 <10^{-4},10^{-5.4}$, which is starting 
to reintroduce a hierarchy problem (this time in the bulk).  If we accept
this limit on the symmetric part of the warping, then eqn (\ref{ccbound2} 
gives 
\be\label{ccbound3}
\frac{\rho}{M_5^4} (1+\omega) <\epsilon^{1/4} = 10^{-4},10^{-5.4}\ .\ee
This is actually weaker (even if $1+\omega$ is order unity) 
than eqn (\ref{ccbounds}), but it does use a value
of the bulk energy density that may be undesirably small.

\section{Discussion}\label{s:discussion}

We have seen how the $\ap$ expansion of string theory can communicate 
symmetry breaking effects between different sectors of a compactification
model -- in this case, Lorentz violation is transmitted from gravitation to
the standard model particles.  To our knowledge, this is the first use of
$\ap$ corrections for such a purpose (but do note that bounds on 
noncommutativity and Lorentz violation in string theory have been studied 
\cite{Freese:2002yp}).  We
then used the Lorentz violating standard model propagators to put limits
on asymmetric warping in string compactifications.  These limits, in our
opinion, reduce the appeal of asymmetrically warped models of cosmological
constant self-tuning because they require some other mechanism to
reduce the brane energy density to acceptable values of Lorentz violation.
We should note again that our calculations are not complete even to
$\mathcal{O}(\ap)^2$ because the additional terms in the action due to 
the background supergravity field strengths are unknown; however, they are
very generally applicable because they only depend on the warp factors.
In addition, our limits seem somewhat stronger than the bounds derived from
field theoretic loops \cite{Burgess:2002tb}; they improve on previous
limits for the speed of gravity in the interesting region of parameter
space.  We should also note that the limits from the $\ap$ expansion are
good even if we use a relatively weak bound for standard model Lorentz
violation.  The key parameter in our limits is the ratio of the Lorentz
invariant warp factor derivative to the fundamental Planck scale, as opposed
to the Planck scale itself for loops.  From naturalness considerations,
we expect that the ratio will be near unity.

To close, we comment that $\ap$ corrections to the D-brane action seem 
very applicable to braneworld models.  They may be useful in correcting
the backgrounds used in the models, just as the $\ap$ corrections for the
bulk action have been (see, for example, 
\cite{Becker:2002nn})\footnote{Also, in Lorentz violating models, 
nonrenormalizable terms may be needed to maintain causality
\cite{Kostelecky:2000mm}.}.  
They could
also be used, as in this paper, to discuss the communication of symmetry
breaking between brane and bulk fields.  As one particular example, string
compactifications with 3-form fluxes and a no-scale supergravity 
interpretation, which have been of great interest recently, exhibit 
sequestering of supersymmetry breaking \cite{DeWolfe:2002nn}.  That is,
the brane fields do not feel the bulk breaking of supersymmetry at tree 
level.  Although a better knowledge of $\ap$ corrections involving 
bulk field strengths and worldvolume fermions
will be needed, it would be interesting to carry out an explicit calculation
of supersymmetry breaking on the brane due to the 3-forms.  As we have
seen in the case of $SO(3,1)$ breaking, the $\ap$ corrections can be just
as important as loop corrections.  They are also somewhat easier to 
compute, since the known corrections can be evaluated using only 
differentiation and algebra and do not require justifying a regularization
procedure for perturbative gravity.  We hope, therefore, that we have
introduced a new technology for the investigation of braneworld 
compactifications of string theory.

\acknowledgments
I would like to thank M. Gra\~na and P. Bin\'etruy for discussions and
especially J. Polchinski for many helpful conversations and
for reading a draft of this work.  The work of A.F. is supported by 
National Science Foundation grant PHY00-98395.

\appendix
\section{Calculation of Scalar Kinetic Term Corrections}\label{a:calc}

We collect here some of the details of the calculations in section
\ref{ss:ap2}, specifically the Lorentz violating scalar kinetic terms of
\ref{sss:mavs}.  For notational convenience, we will call all spacetime
coordinates $X$, whether they are the compact space coordinates $y^m$ or 
not.

We start by specifying the tangent to the D-brane, which
we take as $\del_\alpha X^\mu = \delta^\mu_\alpha$ with perturbatively
small $\del_\alpha X^m$.  Then the normal bundle vielbein satisfies
\be\label{normviel}
\del_\alpha X^M \xi^{\hat a}_M = 0\ \mathnormal{or}\ \xi^{\hat a}_\mu
= -\delta^\alpha_\mu \del_\alpha X^m \xi^{\hat a}_m \ee
which has solution to second order
\be\label{normviel2}
\xi^{\hat a}_\mu = -\delta^\alpha_\mu \del_\alpha X^m \tilde e_m^{\hat m}
\delta^{\hat a}_{\hat m}\ ,\ \xi^{\hat a}_m = \delta^{\hat a}_{\hat m}
\tilde e^{\hat m}_m - \frac{1}{2} g^{\mu\nu}\delta_{\mu}^\alpha 
\delta_\nu^\beta \del_\alpha X^p \del_\beta X^n \hat e_n^{\hat n}
\delta_{\hat n}^{\hat a} \tilde g_{mp}\ . \ee
Here $\tilde e^{\hat m}_m$ is the vielbein of $\tilde g_{mn}$ and is trivial
at the position of the brane.  Since its expansion would just give powers of
$X^m$ and not derivatives, we will ignore it from now on.  Also, for 
notational convenience, since the normal bundle indices $\hat a$ always
enter through Kronecker deltas with compact space indices, we will abuse
notation slightly and replace $\hat a$ with $m$ in future formulae.

The nonvanishing Christoffel symbols are
\be\label{christoffel}
\Gamma^0_{0m}=\frac{1}{2}\del_m A\ ,\ \Gamma^m_{00}= \frac{1}{2}
\tilde g^{mn} e^A \del_n A\ ,\ \Gamma^i_{jm} = \frac{1}{2}\del_m B\delta^i_j\
,\ \Gamma^m_{ij}= -\frac{1}{2}\tilde g^{mn}e^B\del_n B \delta_{ij}\ee
on the spacetime and
\bea
(\Gamma_T)^0_{00}&=&\frac{1}{2}\del_0 X^m \del_m A\ ,\ (\Gamma_T)^0_{0i}=
\frac{1}{2}\del_i X^m \del_m A\ ,\ (\Gamma_T)^i_{00}=\frac{1}{2}e^{A-B}
\delta^{ij} \del_j X^m\del_m A\nonumber\\
(\Gamma_T)^0_{ij} &=& \frac{1}{2}e^{B-A}\delta_{ij}\del_0 X^m \del_m B\ ,\
(\Gamma_T)^i_{0j}=\frac{1}{2}\delta^i_j \del_0X^m \del_m B\nonumber\\
(\Gamma_T)^i_{jk} &=& \frac{1}{2} \left(\delta^i_k \del_j X^m+\delta^i_j
\del_k X^m - \delta_{jk}\delta^{il}\del_l X^m\right)\del_m B
\label{chrisT}
\eea
on the tangent space.  This is just the Christoffel symbol of the pulled-back
metric $g_{\alpha\beta}$.  
Plugging into eqn (\ref{extrinsic}), we can very 
easily get eqn (\ref{omega}) for the extrinsic curvature.  

Then, from eqn (\ref{riemannt}), the tangent Riemann components out to 
$\mathcal{O}(\del X^m)^2$ are
\bea
(R_T)_{0i0j} &=& -\frac{1}{4}a^mb_m \delta_{ij} +\frac{1}{2}a_m b_n
\delta_{ij}\del_0 X^m \del_0 X^n -\frac{1}{2} a_{(m} b_{n)}\del_i
X^m \del_j X^n \nonumber\\
&&-\frac{1}{2}\left( a_{mn}+\frac{3}{2} a_m a_n\right)
\del_i X^m\del_j X^n +\frac{1}{2}\left(b_{mn}+\frac{1}{2}b_mb_n\right)
\delta_{ij}\del_0 X^m\del_0 X^n\nonumber\\
(R_T)_{ijkl} &=& \frac{1}{4}b^2\left(\delta_{ik}\delta_{jl}-\delta_{il}
\delta_{jk}\right)+\frac{1}{2}\left(b_{mn}+\frac{3}{2}b_m b_n\right)
\nonumber\\
&&\times
\left( \del_i X^m \del_k X^n \delta_{jl} +\del_j X^m \del_l X^n \delta_{ik}
-\del_i X^m \del_l X^n \delta_{jk} -\del_j X^m \del_k X^n \delta_{il}\right)
\ . \label{rtcomp}
\eea
The other components are only $\mathcal{O}(\del X^m)^2$, so they square to
$\mathcal{O}(\del X^m)^4$.  When squaring, we also need to take into 
account the second order part of the inverse metric pull-back.  Note that
the only terms involving $\Lambda$ come from $(R_T)_{0i0j}$.  To get
$(R_T)_{\alpha\beta}$ we just contract the Riemann tensor, being careful of
second order terms in the metric.

From eqn (\ref{riemannn}), the normal bundle Riemann tensor is always second
order in $\del X^m$, so it does not contribute to the kinetic terms.  That
leaves just $\b R^{mn}$.  We end up with
\bea
\b R^{mn} &=& -\frac{1}{2} \left( a^{mn}+a^m a^n\right) -\frac{3}{2}b^{mn}
+\tilde R_{pmnq}\del_\mu X^p\del^\mu X^q -\left( a_p{}^{(m}+
\frac{3}{2}a_pa^{(m}\right)\del_0 X^{n)} \del_0 X^p \nonumber\\
&&-\frac{1}{2}\left(b_p{}^{(m} -\frac{3}{2}b_p b^{(m}\right)
\vec\del X^m\cdot \vec\del X^n
-\frac{1}{2} \left(a^{mn} +\frac{5}{2}a^m a^n\right)\delta_{pq}\del_0 X^p
\del_0 X^q \nonumber\\
&&+\frac{1}{2} \left(b^{mn}-\frac{1}{2}b^m b^n \right)\delta_{pq}
\vec\del X^p\cdot \vec\del X^q\ .\label{rbarcomp}
\eea
Here $\tilde R_{mnpq}$ is the Riemann tensor of $\tilde g_{mn}$; note that
it enters in a Lorentz invariant fashion.
Getting the action (\ref{scalarmav}) is just a matter of squaring.

\bibliographystyle{JHEP-2}\bibliography{lorentz}
\end{document}